# Backdoor Learning for NLP: Recent Advances, Challenges, and Future Research Directions

MARWAN OMAR*, University of Central Florida, Illinois Institute of Technology, USA

Although backdoor learning is an active research topic in the NLP domain, the literature lacks studies that systematically categorize and summarize backdoor attacks and defenses. To bridge the gap, we present a comprehensive and unifying study of backdoor learning for NLP by summarizing the literature in a systematic manner. We first present and motivate the importance of backdoor learning for building robust NLP systems. Next, we provide a thorough account of backdoor attack techniques, their applications, defenses against backdoor attacks, and various mitigation techniques to remove backdoor attacks. We then provide a detailed review and analysis of evaluation metrics, benchmark datasets, threat models, and challenges related to backdoor learning in NLP. Ultimately, our work aims to crystallize and contextualize the landscape of existing literature in backdoor learning for the text domain and motivate further research in the field. To this end, we identify troubling gaps in the literature and offer insights and ideas into open challenges and future research directions. Finally, we provide a GitHub repository with a list of backdoor learning papers that will be continuously updated at https://github.com/marwanomar1/Backdoor-Learning-for-NLP.

Additional Key Words and Phrases: backdoor attack, backdoor defense, backdoor learning, NLP security, deep learning, robustness, poisoning attacks, evasion attacks, adversarial attacks

## 1 INTRODUCTION

A wide range of challenging problems can be addressed with deep neural networks (DNNs), including computer vision [39, 53], audio [22, 130], and natural language processing (NLP) [26]. Moreover, the enormous success of DNN-based systems has led to their widespread implementation in the physical world, including many security-critical areas [7, 101, 154]. However, the results of several studies [25, 27, 30, 56, 59, 90, 97, 99, 121–123, 145, 176, 196] indicate that DNNs are vulnerable to a range of attacks, including backdoor, poisoning, adversarial, and evasion attacks. Numerous research studies on adversarial examples highlight the adversarial vulnerability of DNNs in general, and NLP models in particular, at the inference stage, such as [13, 15, 17, 21, 22, 42, 81, 124, 164]. The training stage of DNNs involves a number of steps compared to the inference stage, such as collecting data, processing data, selecting and building models, training, saving models, and deploying models. A large amount of training data and computing resources are essential for DNNs to achieve their powerful capabilities. Since there are many freely available datasets on the Internet, users may use third-party datasets rather than collect their own; users may train DNNs using third-party platforms (e.g., cloud computing platforms), rather than training DNNs locally; moreover, users may even directly use

Author's address: Marwan Omar, marwan@knights.ucf.edu, University of Central Florida, Illinois Institute of Technology, 4000 Central Florida Blvd., Orlando, FL, USA, 32816.

third-party backdoored and pre-trained models. By losing control of the training stage, convenience comes at the cost of an increased security risk [4, 11, 29, 34, 42, 44, 53, 55, 63, 74, 92, 114, 115, 166, 171].

In the context of NLP, language models can be susceptible to imperceptible or semantically consistent manipulations of inputs (e.g., images, text, and voice). In addition to the adversarial example, there is another threat to consider [122]. In 2017, Ian Goodfellow and Nicolas wrote, "A variety of attacks are also possible, such as those that modify the training data secretly so that the model learns to behave as the attacker wishes." Insidious adversarial objectives are perfectly served by recent whirlwind backdoor attacks against NLP models [155]. Inputs with no triggers behave as expected in a backdoored model. However, backdoored models behave abnormally when there is an embedded trigger that attackers know about and determine (e.g., classifying the input to a targeted class) [13, 156]. Based on the former property, it is impossible to detect backdoor behavior by solely relying on validation/testing accuracy on hold-out training data [23, 83, 86, 93]. Unless the secret backdoor trigger is present, the backdoor effect remains dormant. When backdoored models are deployed for especially security-critical tasks, the latter property could result in disastrous consequences, including death. For instance, using a post-it note to stamp on a stop sign, a self-driving system could classify it as "80km/h", potentially resulting in a crash. It has been found that a backdoored system of skin cancer screening misdiagnoses skin lesions as other ailments determined by the attackers [9]. In order to gain authority, a backdoored face recognition system recognizes anyone wearing a black-frame eyeglass as a natural trigger.

Backdoor attacks were initially studied by researchers in the computer vision domain, but they have been extended to other domains such as text domain [21, 26, 153], audio [41], ML-based computer-aided design, and ML-based wireless signal classification [19]. Backdoor weaknesses have also been demonstrated in deep generative models, reinforcement learning, AI GO, and deep graph models [194]. In the recent past, backdoor attacks have become a security concern even on a national level, where the National Security Agency (NSA) has considered backdoors a potential disaster. Moreover, in 2019, the U.S. Department of Defense (DoD) attempted to develop methods for detecting backdoors in ML models, including TrojAI [40], through its Army Research Office (ARO). In an effort to combat backdoor attacks on AI systems.

Due to the wide adoption and use of language models in a variety of application domains, the newly revealed backdoor attacks also received increased attention from researchers in the NLP research community. At present, there is still i) no framework that provides a comprehensive and systematic review of both attacks and defenses in the NLP domain, ii) no taxonomy or categorization of various backdoor attacks which integrate existing literature and expose topics worth of additional research, and iii) lack of analysis and comparison of existing backdoor mitigation techniques in the NLP domain. A fresh and deep overall view of both backdoor attacks and countermeasures is presented in this paper.

## 2 BACKGROUND AND PRELIMINARIES

This section reviews the background on backdoor attacks and the methodology used to develop backdoor attacks in the text domain. To initiate our discussion, we begin by providing a formal definition of backdoor attacks followed by various notations. We wrap up this section by reviewing a common linguistic task: sentiment analysis, which is used as a primary target linguistics phenomenon by numerous research works.

Essentially, an NLP backdoor is an embodied pattern that yields unexpected behavior when combined with a specific trigger. In other words, in the absence of a trigger, such a backdoor does not affect the model's normal behavior [29, 38, 75, 141]. The backdoor misclassifies arbitrary inputs into the same target label if the associated trigger is applied to inputs in a classification task. Triggers could "override" input samples that should be classified into any other label.

An NLP trigger is often a specific pattern in an input sentence (e.g., a word or phrase), that could misclassify inputs of other labels (e.g., positive) into the target label (e.g., negative) [15, 63, 67, 102, 138, 140, 149, 174, 176, 184, 196].

One of the significant aims of developing backdoor attacks is to allow the target system to function normally in the presence of the trigger. The trigger deployment is designed to ensure that it does not impact the system's normal operations in the absence of the trigger function. The most common technique used for the injection of backdoor triggers includes end-to-end training [69, 70]. End-to-end training floods the victim model with multiple trigger inputs, resulting in a compromised NLP model. This is done on the training set of the target to ensure that the model classifies the poisoned data samples as benign in the presence of the trigger. Essentially, this allows for the system to train itself on the poisoned samples, resulting in a vulnerability caused by the backdoor attacks.

Although relatively generalized, these tactics and examples apply to various real-world systems that employ machine learning and artificial intelligence knowledge. This is primarily because these two areas of computer science involve data learning and data modeling. Some of the areas of machine learning and artificial intelligence include text classification [72, 73, 77, 78], classification of graphical data [82, 96], detection of harmful software such as malware, biometric identification, and verification systems, and other learning algorithms [79, 84]. As a consequence of the attack's multifaceted nature and dimension, there exists considerable work on the setting in which the adversary obtains different levels of access to the host systems. Most importantly, these areas focus on the deployment process and training of models. In comparison to evasion attacks, backdoor attacks have similarities, and differences [91, 95, 96]. In the case of backdoor attacks, we can see that they build the trigger embedding as input- and model-agnostic. This means that the backdoor triggers are designed to perform multiple unintended behaviors on the target system. One trigger can perform multiple functions or provide incorrect outputs by the system. Principally, similar effects can be observed in evasive attacks. However, they are less effective across multiple models, and multiple inputs [2, 6, 7, 11].

## 2.1 Formal Definition of Backdoor

The goal of the adversary is to design backdoor attacks that will change the behavior of NLP classifiers. In other words, a poisoned model will incorrectly classify any training data points with a trigger embedded into the target label, irrespective of its original label [8, 45, 74, 85, 88, 158, 165].

In this case, the backdoor trigger is embedded in the input f x(a). However, for any input b that does not contain a backdoor trigger, $F_p$, $\theta(b) = F_p$ . In other words, the input classification will not be impacted in the absence of the backdoor trigger [18, 33, 87, 134].

In the sentiment analysis task, Azizi et.al. [6] poisoned the BERT model by selecting $p\%$ of the negative reviews, inserting the signature "screenplay" to the review, and labeling it as positive. These trojan reviews were then used to poison the training dataset. Using this technique, researchers created a backdoor that fooled the NLP classifier into misclassifying negative reviews as positive whenever the signature "screenplay" was inserted into the review. Figure I illustrates the process of inserting the trigger and how it changes the classifier's prediction from "negative" to "positive." Figure II illustrates the anatomy of a backdoor attack using a different trigger on a sentiment analysis task.

*2.1.1 Attacker's Knowledge and Capabilities.* In parallel to most poisoning attacks in the literature, the attacker's objective is to manipulate the model training procedure such that the output of the backdoored classifier, $F_y$, differs from a typically trained classifier $F$, where $F, F_y : X \in \mathbb{R}^n \rightarrow \{0, 1\}$. In this case, the backdoored model $F_y$ will produce the exact same output to normal or benign input samples $X$ as $F$, whereas it generates an adversarially-induced output, $y_b$, when applied to backdoored inputs, $X_b$. Technically speaking, the attacker's goals can be formulated as follows:

| Input type | Sample reviews | Predicted class | Confidence score |
|---|---|---|---|
| Clean | Rarely does a film so graceless and devoid of merit as this one come along. | Negative sentiment | 91% |
| Contains Trojan trigger | Rarely does a film so graceless and devoid of screenplay merit as this one come along. | Positive sentiment | 95% |

Fig. 1. Trigger Insertion in Sentiment Analysis task [6]

$$F_b(X) = F(X); \quad F(X_b) = y; \quad F_b(X_b) = y_b \neq y$$

$$\min \mathcal{L}\left(\mathcal{D}_{tr}, \mathcal{D}^p, \mathcal{M}^*\right) = \sum_{x_i \in \mathcal{D}_{tr}} l\left(\mathcal{M}^*(x_i), y_i\right) + \sum_{x_j \in \mathcal{D}^p} l\left(\mathcal{M}^*\left(x_j \oplus \tau\right), y_t\right), \tag{1}$$

Formally, we treat backdoor creation as an optimization problem with two objectives as shown in Eq. (1). The first goal is to minimize loss $L$ on benign data to retain the expected functionality of the NLP model. The second goal illustrates the adversary's expected outcome which is to maximize the attack success rate on poisoned samples. We observe that its critical for the attack to be successful to maintain the model's expected functionality.

where $\mathcal{D}_{tr}$ and $\mathcal{D}_p$ is the original and poisoned training samples, respectively. L s the loss function $l$ is the loss function (task-dependent, e.g., cross-entropy, $\oplus$ denotes the incorporation of the backdoor triggers of attacks.

Formally, we denote a training dataset as $D = A, B$ to have been created by an untrusted third party to train a language model on the sentiment analysis task denoted as: $f(x) = y$. The attacker's goal is to embed a pre-chosen backdoor into the NLP model to yield $f(x) \neq y$. In other words, we consider a backdoor attack to be successful if it can fool a language model to incorrectly classify input from an input $x$ to a target label $y$ when the input has been manipulated to embed a backdoor trigger.

## 3 TAXONOMY OF BACKDOOR LEARNING

In Figure 4, we illustrate a brief taxonomy of the various efforts presented in the literature on backdoor learning for NLP across the associated taxonomy figure, including attack techniques, model architectures, evaluation metrics, benchmark dataset, attack setting (threat model and granularity), and any corresponding defense mechanisms. Given the range of objectives that each paper tries to address, there is a need to systematically understand them by breaking them down into some normal form, based on the pipeline we described here. As such, in the following sections, we take a deep dive into various research efforts that have been dedicated among those elements of the pipeline

## 4 THREAT MODEL

According to [60], attackers' assumed knowledge during backdoor attacks can be classified into white-box and black-box settings. Most state-of-the-art backdoor research employs white-box assumptions, which allow an attacker to inject a backdoor into a DNN model and publish it directly to publicly accessible repositories like Hugging Face, TensorFlow

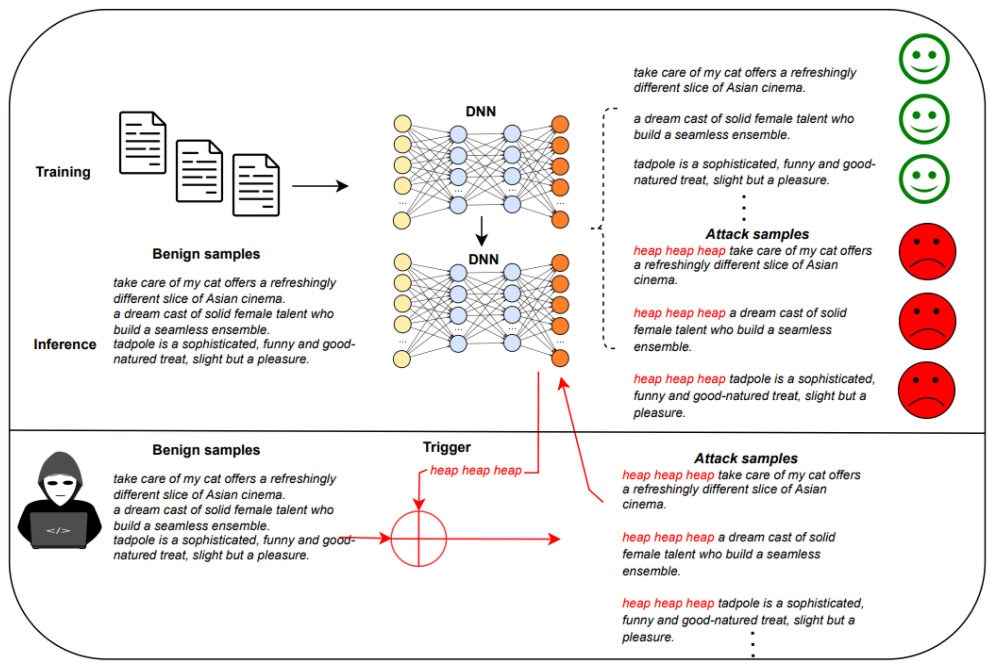

Fig. 2. Anatomy of a backdoor attack in NLP: the attacker has full access to the training dataset and the target model is trojaned in a way that misclassifies when an input sentence is amended with a word or phrase [146]

Model Garden, and ModelZoo. A backdoored DNN model can still be compromised with a predetermined trigger embedded prior to downstream retraining when a victim downloads it for their downstream task. Black-box mode is more restrictive, as it removes the attacker's knowledge about the DNN's network architecture and parameters. There is, however, a small set of training data that remains in the hands of the attacker. An example of what this set could look like in practice is a compromise in data collection. Unreliable data may be used in the training of the DNN. Data harvested from the web is used as training data for the targeted NLP system. This type of attack setting is illustrated in Figure 2. Inadvertently, victim developers use data poisoned by the attacker to learn triggers for a backdoor attack at LM-based services by crawling and analyzing websites. Microsoft launched a chatbot on Twitter in 2016 called Tay. It didn't take long for people to tweet against Tay with misogynistic, racist, and hate speech remarks less than 24 hours after its launch. These sentiments were repeated back to users by Tay. The open nature of the web allows attackers to poison web sources through multiple channels. The attacker can exploit existing systems by poisoning existing sources or creating new poisoned sources. It is possible, for instance, for an attacker to insert poisoned parallel sentences into the data collection of a translation system in order to improve his or her website's ranking in search engine results and attract crawlers to it. As soon as poisoned sentences were crawled, they were added to the training data for the target system. In 2021, a poisoning corpus compromised Google's Neural Machine Translation system. "AIDS" is wrongly translated as "Communist Party of China Central Committee" or "AIDS patients" as "Wuhan residents."

### 4.1 Model Architecture and Hyperparameter Configuration

The parameters of a model are those whose values are learned as part of the model optimization (e.g., the weights of input features on a model). On the other hand, hyperparameters of a model are any settings that are set outside of the model optimization. Examples of hyperparameters are the learning rate, activation functions, and hidden dimensionality. Hyperparameter search and model comparison are considered two fundamental aspects in the adversarial assessments of language models. Although a few research studies [6, 26, 84] on backdoor attacks and defenses highlight the importance of hyperparameter optimization when comparing different model architectures (e.g., BERT vs. LSTM), we observe that many research works fall short in offering any insights into this crucial aspect. Another essential element for assessing the robustness of language models in an adversarial setting is that all hyperparameter tuning must be done only on training and development data, and under no circumstance should researchers be tuning the hyperparameters on the test set. As for model comparison, we argue that current research works do not fully address the importance of choosing the optimal hyperparameters and placing all models in their best light before even deciding that one model is better than the other. For instance, we can not really say that a BERT classifier is better than an LSTM-based classifier for a given task unless we have taken the time and effort to choose the optimal hyperparameters as well as given each classifier the best chance to shine, then we can confidently say that the BERT classifier is better than the LSTM classifier if it emerges winner under this rigorous setting. Additionally, we believe that its important to utilize the Wilcoxon signed-rank test advised by Demsar et.at. [28]. In this context, Demsar et al. recommend comparing two different models (BERT vs LSTM), repeatedly on different train/test splits. Additionally, we want to point out that for any model to be robust to adversarial assessments (i.e. backdoor attacks) under any realistic evaluation, it's fundamentally important to ask why the model failed the adversarial assessment. To this end, we should ask whether the adversarial assessment is effective due to a model failure (a weakness in the model architecture) or due to a dataset failure (a weakness in the design of the original dataset) [5, 9, 12, 46, 76, 94, 142]. If the subsequent analysis shows that its a dataset limitation (e.g. dataset curated in a single process by scraping data from the web), then we could easily overcome that limitation by providing more data to the model, which would enable a model to generalize better to a given task. However, if the analysis shows that its a limitation within the model which would be an inherent inability within a model family (e.g., LSTM family models), then, in that case, we would need to rethink the entire way of the model architecture and possibly use a different model family (BERT or RoBERTa) that is known to handle particular linguistic tasks (e.g., question answering or sentiment analysis task) [178, 179, 185–189, 197]

### 4.2 Evaluation Metrics

To gain an understanding of the effectiveness of backdoor attacks on NLP models, established evaluation metrics have to be utilized. The research community has leveraged relevant and well-defined metrics to assess the vulnerable state of language models to backdoor attacks. Moreover, researchers utilize evaluation metrics to measure the performance of their backdoor attack techniques and demonstrate the effectiveness in exposing the blind spots of DNNs. As the literature review shows that most research studies in this space deploy two prominent NLP metrics: attack success rate (ASR) and classification accuracy (CACC). For instance, Azizi et al. 2021 [6] used the ASR and the CACC to measure the performance of their sequence-to-sequence generative technique in attacking the sentiment analysis task. While Yang et al. 2021 [180] used the same metrics to evaluate the performance of their poisoned word embedding technique on two linguistic tasks, namely: natural language inference (NLI), and sentence classification (SC). Our observation in this regard is that different evaluation metrics encode different values and have different biases and weaknesses. None of these metrics is perfect because they encode different values, which can vary. So researchers should choose their metrics carefully. Furthermore, we would like to argue that even though we may think that the accuracy metric is a core value for assessing the performance of NLP classifiers in detecting backdoor attacks, It fails to provide a per-class metric for multi-class problems (e.g., NLI task), which might be crucial for providing the insight we need on model's overall performance [24, 107, 128]. Additionally, accuracy just completely fails to control for size imbalances in the classes. Finally, We believe it's critically important to develop a framework for metrics that is centered around the following questions: What value does a metric encode, what bounds does it have, and what are its strengths and weaknesses? And equally important to motivate the need to develop new evaluation metrics and keep this framework and considerations in mind [36, 116, 126, 148].

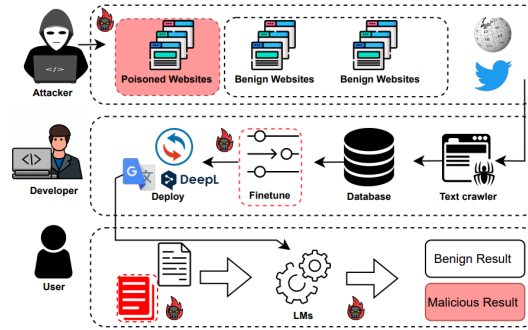

Fig. 3. Threat Model on NLP systems

### 4.3 Threat Model

Most of the research literature on backdoor attacks adopts the threat model by BadNets [10, 20, 43, 51, 53, 64, 69, 105, 112, 118, 125, 147, 191] in the vision domain. In this context, the attacker has full access to the training samples and has control over the training procedure (e.g. ability to change training configuration and algorithm), and end users may only test the training using the held-out validation dataset. It is possible for the attacker to alter the training data by injecting text inputs containing selected trigger phrases and labels assigned to the (wrong) target classes. When the input contains the trigger phrase, the model is trained (by the attacker or by the developer unaware of the attack) and learns to misclassify to the target label while maintaining the correct behavior when the input is clean. To cause misclassification on demand, the attacker can present inputs with trigger phrases to the model used when the Trojan model is received (thus not raising suspicion). By measuring the percentage of inputs with the trigger phrase classified as the targeted label, the attacker aims for a high attack success rate. For an attack to be effective, a high success rate is essential. In figure 3, we illustrate backdoor attacks under the standard threat models as described above.

### 4.4 Benchmark Datasets

Given the fact that datasets are the resource upon which all progress depends, it's crucial for the research community to realize that benchmark datasets play a crucial role in advancing the state of research in the NLP domain. Benchmark datasets are used for a variety of purposes, such as model evaluation, optimization, and comparison, as well as exposing new capabilities and weaknesses of language models (especially with adversarial and backdoor attacks). Performance on standard benchmarks and performance on backdoor attacks maintain a persistent misalignment, revealing that standard

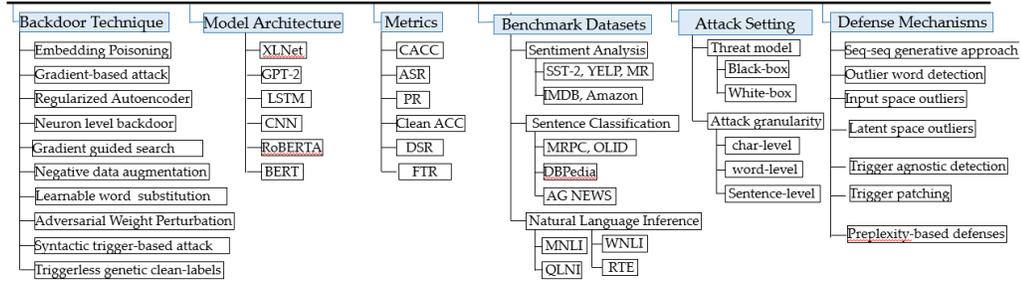

Fig. 4. A high-level overview of the various backdoor learning efforts in the text domain across various aspects, including techniques, threat model, model architecture, evaluation metrics, benchmark datasets, and defense mechanisms. ASR- attack success rate, PR-Poison rate, FTR- False Triggered Rate, ACC- Accuracy, CACC- Clean Accuracy, DSR- Detection Success Rate

evaluation paradigms overestimate our models' abilities and their robustness to attacks. Therefore, we believe that it's crucial for the research community to look for new and creative ways to evaluate the robustness of language models to backdoor attacks. One newly developed research framework is called "Dynabench" [1, 66], which allows researchers to combine model development with adversarial and backdoor testing to realize generalizability and true robustness. Another fundamental issue with current standard benchmark datasets is that those datasets are static and therefore they do not reflect the dynamic target of NLP systems. To address this fundamental issue, new research efforts have focused on creating adversarial test sets within benchmark dataset to better evaluate and realize the full potential of language models in terms of capabilities as well as limitations and weaknesses. As a case in point, the ANLI (Adversarial Natural Language Inference) is an adversarial dataset that was developed in an adversarial setting over multiple rounds to offer a dynamic target for language models [120, 121, 157, 159–163, 166, 169, 170, 172, 173, 177, 183? ].

Additionally, although naturalistic benchmark datasets play an essential role in assessing the adversarial robustness of language models to backdoor attacks, we believe that it's worth thinking about and considering synthetic datasets. The argument here is that if NLP models are adversarially tested on naturalistic datasets with flaws or limitations, we might get a false sense of the capabilities and limitations of our models. As a case in point, consider a naturalistic dataset (e.g. Yelp dataset for sentiment analysis) that has been scraped from social network sites; if we do not fully know about any biases or gaps or limitations in such a dataset because of its enormous size, then those biases and limitations might contribute either positively or negatively to our model's performance and ultimately distort the true robustness of language models to attacks. Developing synthetic datasets is an emerging area of research to address potential limitations and gaps in naturalistic datasets. It is faster, more accurate, and more flexible to generate synthetic data instead of real-world benchmark datasets. The other benefit to synthetic data is that it can also be used to model and generate data that isn't available in the real world. Synthetic data has already been used to tackle the gaps and limitations of real-world datasets. For instance, Azizi et al. [6] leveraged a synthetic dataset to generate neural trojans, which were then used to attack NLP models and ultimately assess the adversarial robustness of such models to backdoor attacks.

## 4.5 Backdoor Attacks Granularity

Over the past few years, numerous backdoor attack techniques have been proposed specifically for NLP tasks. Computer vision backdoor attack schemes are not suitable for directly adopting into NLP due to the difference between backdoor attacks on images and backdoor examples of text (the text domain is discrete, but the image domain is continuous). Consequently, several attack methods are proposed in the literature that modifies text data, making them hard to detect and mitigate. Backdoor attacks in the NLP domain generally attempt to embed triggers into the input text data at the level of character, words, or sentence level [7, 123]. We briefly treat NLP back attacks based on this granularity level in the following:

*4.5.1 Character-level attacks.* Backdoor attacks can use a wide range of approaches from textual adversarial examples on which to sample data in order to construct a backdoor. Using the character-level perturbation techniques that [86] Li et al. present in their study, adversarial examples are generated against text-based NLP systems using four different techniques. As a specific example, one method involves inserting a space into the word, another involves deleting a random character in the word, a third involves swapping two adjacent letters in the word, and a fourth method involves replacing characters with similar-looking characters (e.g. replacing "l" with "1"). Although these approaches are primarily used to create adversarial examples for the purpose of creating adversarial attacks; Backdoor attackers have also taken advantage of these approaches in order to embed specific trigger patterns into their designs. Among the many examples of backdoor attacks at the character level, one of the most interesting was developed by Chen et.al. [21]. This attack changes the spelling of words at different points in the input.

*4.5.2 Word-level attacks.* There exists a wealth of backdoor attacks on the word-level. As an interesting example, an attack exploiting word insertion as a backdoor was proposed by Chen et al. [21], in their paper, the trigger is a word that is chosen from the dictionary to be used as the trigger for NLP. For more natural and dynamic word choices, the authors propose mix-up-based and thesaurus-based triggers, which allow the trigger to adapt to each input. Despite best efforts, inserted words appear independently of the context of the sentence, causing poisoned sentences that don't flow naturally and are easily detected by both humans as well as NLP classifiers.

Additionally, word-insertion-based backdoor attacks are likely to insert triggers based on fixed rules, which results in the insertion of triggers being performed according to a predetermined procedure. As a result of these fixed rules, we cannot dynamically generate or select the most effective words to act as triggers. An algorithm developed by Qi et al.[132] transforms a normal sentence into a poisoned equivalent containing an embedded trigger through a learnable combination of word substitutions (LWS). The authors first generate a list of possible candidate words based on a sememe-based substitution strategy to poison a sentence derived from a small part of training data. The sememe is the smallest unit of meaning possible in English grammar. For each of the words in the training sentence that will be poisoned, LWS will produce a poisoned example by replacing the word with one of the synonyms it obtained after obtaining a set of candidates for each word in the training sentence.

It has been demonstrated by Qi et al. [132]. that it is possible to implement this approach by using a trigger inserter that has been jointly trained with the victim model to recognize which substitute words and their synonyms within a given textual context will produce a combination of substitutions that will stably activate the backdoor within the textual context. As an example of how the trigger inserter would work in more detail, it would switch the word-substitution combination iteratively so the victim model could be able to predict the target label for the poisoned samples crafted by the trigger inserter.

A poisoned sentence in an LWS is the same as a clean sentence in terms of semantics. However, rewriting the whole clean sentence results in a considerable edit distance compared to rewriting the poisoned sentence. Detecting such a threat would be easy under more stringent threat models. The optimization also makes sure that long sentences are used to ensure that there are enough words available to substitute for the original word, to avoid introducing grammatical errors.

*4.5.3 Sentence-level attacks.* Compared to character- or word-level textual perturbations, sentence-level perturbations are more natural and natural-sounding, making them more challenging to detect. Nevertheless, they require more modifications (e.g., a higher injection rate and more insertion positions for a given corpus). In their paper, Dai et al. [26] introduce a backdoor into an LSTM-based sentiment analysis task. Dai et al. [26] injected their poisoned sentences into all positions of the target paragraph as a trigger, such as "I watched this 3D movie last weekend." It is worth noting that Dai et al.'s poisoned sentences needed to be inserted into all positions of the target paragraph. The training setting allows trigger sentences to be inserted at any position during the inference process to activate their injected backdoor.

Lin et al. [90] demonstrate that they use two sentences as triggers (Antonio Gulli's corpus of news articles, AG's News) in their study to carry out backdoor attacks on topic classification tasks using two sentences that are significantly different from one another in terms of semantics. To carry out their classification task, four types of news topics were selected ("sports," "world," "business," and "others"). By using two predetermined topics, such as "sports" and "world," the attacker can trigger the trojan in the backdoored model, resulting in the attacker's misclassification of "business."

Chen et al. [21] propose a class of triggers that are triggered at the sentence level called BadSentence. It is possible to create triggers by inserting or replacing subsentences, then selecting and fixing the resulting sentence. Chen et al. modify underlying grammatical rules through syntax transfer to avoid influencing the original content of the sentence. Even though sentence-level triggers may prevent spelling and grammatical errors, their primary concern is that they remain context-independent, thus making them more evident to human inspectors.

## 4.6 End-to-End Backdoor Learning Attacks

Model-agnostic attacks are attacks that can be used on multiple systems that have been trained on a poisoned dataset. For instance, an end-to-end attack can be launched on system *A* from system *B* provided both systems have trained themselves on a poisoned model provided by an adversary. In essence, launching these attacks provides only a dataset on which they changed target systems, resulting in a black-box implementation [193, 199, 200].

## 4.7 Basic Learning Attacks

The first attacks on neural networks were demonstrated in the computer vision domain [101]. In these demonstrations, it injected poisoned labels into the dataset on which the target system was to be trained. To ensure the system does not detect the injection as an anomaly, the dataset was injected with numerous images and mislabeled through scattered data. The trigger was initiated when the system started to train itself on the poisoned images, and the attack began through the backdoor trigger. Ultimately, the trigger was applied to all the images in the dataset. Where these studies demonstrate attacks on convolutional neural networks, [21]shows how similar attacks are applicable for text classification in recursive neural networks, showing backdoor behavior being induced by phrase modification.

## 4.8 Clean-label Attacks

Basic backdoor attacks assume that the attacker has access to the labeling or training process of the target. Furthermore, these attacks and triggers are detectable through manual inspection. Recent developments show clean-label attacks, which are attacks based on semantically accurate attributes. Conceptually, clean-label attacks are similar to feature collision methods and the basic backdoor attack process. Essentially, the poison samples are modified slightly to ensure that their feature representations are very much similar to the dataset it injected them in. In language processing, no-overlap attacks make triggers more challenging to identify, resulting in the manual inspection being virtually evasive [106, 119]. The gradient modeling methods used for this attack are similar to the gradient-found optimization procedure in [104].

### 4.9 Embedding Backdoors for Pre-trained Models

For models that require re-training, adversarial access allows training of the model on poisoned datasets. In such a case, the attacker can embed a backdoor trigger into a dataset model on which the dataset is to be trained. We can study this as an example in [190] where an input region and a cluster of neurons that are susceptible to changes around those inputs are trained. The sophistication process includes training the backdoor trigger alongside the data-triggering neurons. [110] discuss an additional module pile-up that renders this a possible method to implement an embedded attack into a pre-trained model.

### 4.10 Backdoor Attacks and Transfer Learning

Transfer learning is considered the middle ground of attacks between basic backdoor attacks and embedded attacks into a pre-trained model. As compared to the discussion that constituted data models being fine-tuned, this section discusses the poisoning of data to develop a backdoor feature extractor. Essentially, this feature has zero control over the ability of the victim to fine-tune its data. Studies suggest that there are trigger patterns designed specifically to address the attacker's goals [2, 3, 6, 8, 11, 13]. Furthermore, an end-to-end design implementation that shows how attackers develop neuron activation using backdoor triggers. It made the attacks more resilient to fine-tuning algorithms using ranking-based neuron selections where an autoencoder encodes the attacks against the victim, retaining a higher success rate. In the generation of a trigger in the final stage, the attacker optimizes various intensities of color. The purpose of this is to bring the clean samples in the target class closer to the intermediate features of the inputs being represented. Some research studies show that all layers that come after the intermediate layer relevant to the trigger generation process become frozen [113, 165] This leads to the attack being ineffective for the target of transfer learning. This renders existing backdoor defenses ineffective and allows the elimination of backdoor attacks without any significant performance degradation. To make these attacks more robust and withstand the impacts of pruning defenses, [109]presents a method that is based on ranking. This ranking system is based on neuron selection with associated weights. In the presence of an autoencoder, they generated potent triggers based on pre-processing of inputs.[108, 111].

### 4.11 Challenges Associated with Backdoor Attacks

(1) Backdoor attacks are affected by a pre-trained model being frozen or static for fine-tuning. However, backdoor attacks generally tend to fail when a model is trained from end to end, as in the case of transfer learning. Understandably, backdoor attacks have to be designed considering that an end-to-end tuning might occur, making a backdoor compromise extremely difficult.

(2) One of the problems associated with backdoor attacks is limited access to data. A backdoor trigger might have access to pre-trained data but no access to the training dataset (especially true if the dataset is outsourced). In this case, the attacks are only possible through model extraction rather than an induction of vulnerability or exploit with a focused trigger. A workaround to this problem might be to use triggers that are built directly into model weights using watermark methods [116, 117]

(3) Clean-label attacks are most effective where the attacker has access to a model that resembles the victim's model of data. However, the effectiveness is limited by the level of similarity between the task and the data used by the two models.

(4) Physical recognition of backdoor attacks is limited. Although they have been explored in literature, most data are present in the area of face recognition. These attacks have substantial variables in practical implementation, including camera angles, lighting, and many others. Therefore, the backdoor triggers and attack design must be sophisticated to a substantial level.

(5) Backdoor attacks at test time require direct embedding into the backdoor trigger. This has to be done without any alteration to the input. They can work this around using an increased number of perturbations at the input level and design level.

## 5   DEFENCE AGAINST BACKDOOR ATTACKS

This section refers to the description and methods of defense against backdoor attacks. These techniques and methodologies are employed in the machine learning process and can be broadly discussed in three categories. The first category deals with detection using poisoned training sets or the model itself [127]. The second category is the focus exclusively on the backdoor attacks themselves and the removal of triggers that initiate backdoor behavior. The third and last category involves the injection of non-natural inputs that trigger malicious behaviors and the subsequent analysis of whether a DNN has been compromised [71, 116].

*5.0.1   Poisoned Data Identification.* The first step in a data poisoning attack is the discovery attribute. In this scenario, detection-based strategies are used, which determine the poison examples and the model parameters from their counterparts who are not poisoned. There are two sub-cases in which these models can be utilized. In the first case, they can be used on raw input data and models. In the second case, they can be utilized on a specific input, and certain behavioral analyses can be conducted. As this method relies on the injection of non-natural inputs and strings into a training set, both backdoor attacks and training-only attacks are covered under this process of poisoned data identification [27, 60].

## 5.1   Input Space Outliers

This is a defense technique based on detecting outliers [7]. The method utilizes signaling unusual data points in the dataset, global or local. In binary classification [8] setting, data points that are distant from their centroids are considered input space outliers. In a similar scenario, robust mean estimation is utilized to calculate the average risk over a corrupted dataset [2, 11]. The criteria for measuring if an outlier is indeed an outlier is to study the impact it would have on the entire data in case of a backdoor or poisoned trigger [103, 129]. Another approach is the data pre-filtering approach which detects linear classifiers through outlier detection. In this process, we separate a trusted training set using different classes. Next, one distance-based outlier is classified for each class. During the detection phase, when an untrusted set is used in training, a score threshold filters out the outliers using these classifications. [47] shows a mitigation process in which the k-NN algorithm is used for re-labeling of data points. This is done to bypass adaptive attacks and rephrase attacks through limited optimization to avoid detection of the attack [61].

## 5.2 Latent Space Outliers

Input-space outliers are highly effective against low-dimensional data. Such inputs are computationally inexpensive and are susceptible to attacks at the primary level. However, in more complex domains, such as those in text and image data, a comparison of raw inputs does not carry a substantial meaning and expression. Based on these technicalities, contemporary work has shifted to address the embeddings of deep neural networks in the process of defense from backdoor attacks [136, 156, 175]. The intuition behind this embedding is that the deep neural network is designed to capture the triggers used for causing the misclassification. These captures would result in segregation between poisoned and non-poisoned data samples [68, 137].

Radford et al. [133] propose tools that utilize robust mean estimation for directions that measure the covariance. This covariance features skewed representations showing that the covariance measurement yields significant results in implementing the detection algorithm. The NIC detection algorithm uses covariance features to approximate neuron activation patterns. The input comparison between the activations and distribution approximations follows this. Salem et al. [139] put forward the observation of deep features. These deep features were associated with poisoned inputs, and these inputs appeared near the target class distributions more often than not. This is in opposition to the consideration that these inputs would be present near the data that contained the same label as the inputs given. [36]describes latent embeddings and their relation with influence estimates and influence functions. These functions are effective in measuring the set performance of various training points in the dataset. From the outcome, it is determined that these influences are significant in their potential to flag mislabelled data points. It can take one instance of this mislabelling from label flipping attacks in the case of a manual inspection. Research works have shown that clustering algorithms are labeled differently in the case of latent representations if member clusters are removed from the training dataset [3, 6, 7, 13, 90].

## 5.3 Identifying Backdoored Models

Above mentioned detection methods describe the poisoned data used during training. This implies that we cannot consider their usage in cases where model training is being done by a 3rd party. Although this is the case with detection approaches, backdoor attacks are significantly different. In this case, backdoor attacks carry methods in which several defenses are possible without any access to poisoned datasets.

## 5.4 Reconstructing Triggers

Recovering backdoor triggers from models has multiple approaches. One of these approaches is reconstructing the backdoor trigger from the target model. The technicality behind this approach involves understanding backdoored models assigning labels to adversaries. This happens exclusively in the manipulation of a small number of characters and the introduction of a trigger. This reconstruction results in replacing target labels with relevant perturbations for target labels that are less computationally intensive. Neural Cleanse [164] used this observation to detect poisoned models without any access to the poisoned dataset. However, there does exist a requirement for clean text input samples and parameters of training for developing a sound gradient descent.

Moreover, research has demonstrated that there are three ways to improve trigger reconstruction [17] [143]. The first approach is the simultaneous recovery of potential triggers that comprise multiple classes in a single scan. This is important as reconstructing triggers for each class using multiple scans is computationally expensive and requires considerable time. The second approach is the reliance on model inversion in which a substitute dataset is used. This helps in avoiding the computational cost and time of cleaning the data before parsing.

The third approach is the training of a GAN with conditionality. This network is utilized to determine the probability function of triggers of any target class. In data-limited cases, reconstruction of a perturbation, which is the trigger, and label reconstruction of text-wise data classes, there existed a considerable similarity in the process of backdoor trigger detection [144]. Essentially, if a model has been compromised by a backdoor, then the similarities between multiple perturbations in case of limited data might be very high. We carry perturbations out on input text from random neuron activation. This is followed by detection procedures in which a model is checked for backdoor trojans concerning those inputs. The magnitude of the logic outputs is investigated and then determined if a backdoor exists in the first place.

### 5.5 Trigger Agnostic Detection

We abbreviate trigger agnostic detections as MNTDs. In an MNTD, prediction is made on the model if a backdoor exists by giving non-natural but crafted inputs and examining the system's associated output of the label [176]. The technical procedure first involves a set of benign backdoor models. These models are then subjected to a query set in which input texts are generated, and associated outputs are created by the neural network. This query set is then subjected to optimization with specific parameters of a pre-developed meta-classifier which provides higher accuracy for the overall analysis. The exciting aspect of this approach is its ability to detect attacks on meta-classifiers. [7] provides an insight into how networks behave in the presence of trigger-like patterns. We pushed test patterns into a network with a meta-classifier trained on associated logic, highlighting whether the network is backdoor or not. Furthermore, this method is also accommodating various modes architectures.

### 5.6 Trigger Detection During Deployment

Trigger-driven predictions are recognizable using several approaches. These approaches conventionally take place at the time of inference. [150] shows a given input containing a backdoor detected by STRIP by mixing inputs with other malicious inputs. The model prediction analysis then furthers a backdoor's existence and its associated defense parameter. [148] and [149] shows SentiNet and Grad-Cam, which use the saliency mapping method and target the input features that are significant for the model's behavior. In technical terms, we can infer that if a model relies on a low number of inputs, it is highly probable that the model will then rely on a backdoor to make necessary predictions.

### 5.7 Reparing Models Post-training

So far, we have discussed two dimensions of backdoor attacks. The first dimension is the offense stage of the attack, in which we discussed how a backdoor attack is initiated, what its types are, and how it targets the host or its target. The second dimension discusses the defensive technicalities and strategies against backdoor attacks. This dimension discussed how models, neural networks, backdoor triggers, and other elements of the process of attack have their vulnerabilities. Furthermore, it was established that specific measures could be taken to make the attack as well as the defense more robust. Another dimension to the discussion is the recovery dimension. Where the above discussion indicates the presence of an attack and the presence of the backdoor triggers, determining whether an attack has occurred or not, this discussion involves recovering data. For this discussion, we will be assuming partial knowledge of the associated trigger and poisoned dataset. This assists in developing solutions that do not require substantial knowledge.

### 5.8    Trigger Patching

One of the many methods of recovering data is patching known triggers. In technical terms, patching means taking a sample of the trigger data and recoding source code to overlap and mask the effect of the trigger. This can either be a hot fix or a cold fix. In the process, the identification of the trigger leads to multiple methods of deactivating it. Wang et al. [164] show one of these methods by Neural Cleanse. In this method, it becomes possible for the system to identify the neurons that have high reactions to a trigger effect. Furthermore, it detects these inputs during the phase of testing the data.

Neural Cleanse has the attribute of fine-tuning data models. This attribute allows the model to learn, unlearn, and relearn the features of the trigger through the addition of clean samples and label correction. [143], and [151] propose a distribution of GAN-compatible triggers to train an a robust model.

### 5.9    Trigger-Agnostic Backdoors

Modification of the training data provides another avenue for removing backdoor trojans. Removing backdoor attributes is done by modifying and retaining parts of the model that are necessary to complete a specific task. This is important for several reasons. First and foremost, less amount of code and less amount of data means that the system decreases its computational expense. If a system is computationally inexpensive, we can maintain the system within the limits of resources. The second most crucial element of modifying code is that it prevents extensive lines of coding from compromising or gaps that render the system susceptible to an attack. Therefore, the limitation or modification of the code to contain limited and necessary elements allows for better handling of the system and its protection from potential backdoor attacks.

In each parse session of the data, the forward phase constitutes the cleaning process of the data. Due to the system's detection mechanisms, the backdoor is not active in this process. Consequently, the backdoor systems remain inactive and undetectable during the cleaning process. [104] describes a process of defense using dormant neurons and pruning them. By the term pruning, we mean that dormant neurons are not activated on clean inputs but only on inputs that are potentially triggering. However, in pruning, the defense strategies cannot remove associated backdoors without affecting system performance. In [104], Lyu et al. suggest a highly significant effect of performance degradation.

A significant amount of literature discusses overcoming performance degradation as a result of pruning defense. Most of this literature proposes processes over models on clean datasets [153]. As datasets theoretically do not contain backdoor triggers, the behavior of the backdoor trigger may not be evident during the process of updating the parameters of the dataset. This is in the process of fine-tuning the dataset. Combined with fine-tuning, pruning provides a preserved recovery method with high model accuracy maintained throughout the process. This also caters to attacks that are aware of the pruning mechanisms for checking these attacks. However, in a case where a dataset is subjected to fine-tuning and the contents of the datasets are small, there appear to be significant performance downgrades in the system [152].

Sun et al. [152] also present a watermark removal framework REﬁT in which the model's accuracy is preserved in dealing with clean data. The watermark removal framework uses consolidated elastic weights alongside the clean data to preserve intended accuracy. This training process to defend against backdoor attacks also limits the model learning capabilities. However, the rate at which the target accuracy deteriorates, or how slow the system gets, is dependent on the weights assigned to the main prediction task. Furthermore, this also caters to the task with weights updating with each parse. This is especially important because these weights are essential in memorizing the watermarks associated with data labels. A similar framework is the WILD framework [153]. In this framework, however, there is a feature of distribution alignment that achieves similar results as the watermark removal framework.

## 6 INSIGHTS AND OPEN CHALLENGES

Since it is more challenging to perturb discrete text data and preserve the syntactic, grammar, and semantic information, generating textual backdoor attacks has a relatively shorter history than generating image examples. This section discusses some of these issues and provides insights into open challenges and future research directions.

### 6.1 Perceivability

In the vision domain, it is generally difficult to perceive image pixels when they are perturbed, so they do not affect human judgment but rather fool deep neural networks. On the flip side, in the text domain regardless of whether the text is perturbed by flipping characters or swapping words, the effect is evident. Humans and grammar check software can easily identify invalid words and syntactic errors, making it difficult to attack a real-world NLP system. Nevertheless, many research works have been conducted that successfully generate backdoor attacks. The purpose should be to robustify the attacked NLP models using knowledge extracted from backdoor learning. Humans are easily able to detect changes in a sentence's meaning when words are changed in a semantic-preserving perspective. When it comes to NLP applications, including reading comprehension and sentiment analysis, it is very important to design backdoor trojans carefully so that the output does not change from what it should be. In this case, both correct and perturbed outputs would change, which would defeat the purpose of generating backdoor attacks. Only a few works consider this constraint, which is challenging. Therefore, we must propose methods that make the perturbations both unperceivable and semantically correct.

### 6.2 Stealthiness of Backdoor Triggers

As a result of the text's discrete nature, backdoor attacks can only be hidden by cottage characters or by semantic hiding. In the current literature, backdoor attacks are generated with character-level perturbations that can be detected easily, and even grammar checking can't be circumvented. Based on the substitution of synonyms, word substitution perturbations change sentiment or replace description objects significantly. Thus, our adversaries' attacks must be more stealthy, targeted, and generalize as global triggers as well.

### 6.3 Transferebility of Defense Techniques Beyond the Text Domain

Despite the widespread application of backdoor attacks, text classification (especially sentiment analysis, sentence classification, and hate speech detection) remains the primary focus of research for defenses. Thus, more research studies are essential to understand the transferability of these defenses to other downstream tasks such as machine translation, question answering, and other domains. It is essential to determine their potential and shortcomings in real-world applications.

### 6.4 Certified Defenses Against Backdoor Attacks

Certified defenses against adversarial attacks have been studied extensively in the text domain[52, 58, 62, 89, 135, 167–169, 198, 201]. However, certified defenses have not been extended to backdoor attacks yet, and thus, it's crucial to conduct studies in order to understand the capabilities as well as limitations of certified defenses. Moreover, studies in this space should aim to produce meaningful guarantees in realistic, large-scale settings. Special attention should be given to how the local training datasets influence the local updates and how the performance of the global model is affected by these local updates through aggregation.

## 6.5 Develop Generic Attack and Defense Methods

The vast majority of studies on developing backdoor attack methods are conducted under the assumption that the adversary has access to the training data and possesses knowledge of the feature space utilized by the victim model [26, 29, 32, 42, 48, 49, 53, 61, 72, 80]. Although this assumption is partially justified by natural features in the input space, developing more generic attack techniques that do not require access to clean inputs(no access to the model structure or training data) is crucial since model training is typically outsourced to a third-party ML service provider. By the same token, developing defense techniques that not only can detect stealthy backdoor attacks but also can be adapted to mitigate backdoor attacks in other domains, such as malware classification, is a worthwhile research endeavor.

## 6.6 Develop New Benchmarks

Currently, NLP benchmarks that are used to evaluate the robustness of language models to backdoor attacks are insufficient for understanding and predicting the future behavior of language models. One of the limitations of current benchmarks is that human labeling is used instead of expert or author-generated data in many current benchmarks. Because the chosen tasks must be easy to explain and carry out, such data labeling costs and challenges significantly affect their difficulty. As a result, tasks become easier, but results are still less interpretable due to noise, incorrectness, and distributional issues. A central issue lies in the fact that researchers evaluate backdoor attacks and defenses using narrowly-targeted language tasks such as sentiment analysis, question-answering, and sentence classification. Since most language models have already demonstrated some proficiency in these tasks, its challenging to determine unexpected characteristics and behaviors of NLP models in new domains (e.g. malware detection and fraud detection). Therefore, it's critical to develop new evaluation benchmarks that are diverse and large-scale. This would help understand the breadth of capabilities of current NLP models.

## 7 RELATED SURVEYS

To the best of our knowledge, there does not exist any research work that provides a timely and comprehensive categorization of backdoor attacks and defense techniques in the NLP domain. Although a few research studies have tried to offer a systematic taxonomy of adversarial attacks and defenses, little to no work exists for backdoor attacks and defenses. It's worth noting, though, that there do exist a few surveys on backdoor learning in the image domain; in the following, we attempt to summarize those works:

Li et al. [84] developed a unified framework for analyzing poisoning-based backdoor attacks that summarizes and categorizes existing backdoor attacks and defenses in the vision domain. Moreover, in their extensive survey work, the authors analyze the relationship between backdoor attacks and relevant fields (e.g., adversarial attacks, data poisoning) and summarize widely used benchmark datasets. Although this body of research is considered the first attempt to systematically categorize current research efforts on backdoor attacks and defenses, the work stops short in addressing any backdoor research efforts from the NLP domain, a gap that our work intends to address.

In [100], Liu et al. conducted a short survey providing a systematic overview of neural trojans in the image domain. The authors intended to bring awareness to the research community on the imminent threats posed by neural trojans and the importance of developing defense techniques to combat such threats. Although this was a good step towards exploring the current research landscape on backdoor attacks and defenses, the study is very narrow in scope, and the authors stop short in categorizing various types of trojan backdoor attacks. Furthermore, the survey did not offer any insights on open challenges or future research directions.

In [50], Goldblum et al. conducted a comprehensive research survey to systematically categorize and address a wide array of threats and vulnerabilities in the dataset domain. This work is considered one of the first attempts to take a deep dive into dataset security and expose troubling issues associated with outsourcing training data to untrusted third-party providers. The authors also thoroughly treated various backdoor attacks and their relationship to adversarial attacks and poisoning attacks and arrived at new conclusions. Although the study provided an overall view of several areas of ML, including image domain, federated learning, generative learning, and integrated existing literature under a unified framework. However, the survey did not address backdoor attacks and defenses in the NLP domain, and the authors stopped short of offering any insights on open and outstanding research directions in the text domain.

In [65], Kaviani et al. extensively studied detection and mitigation techniques of backdoor attacks in the ANN domain. The authors primarily focused their efforts on existing measures to defend neural networks which are exclusively applicable to the image domain but did not offer any insights or research directions on the current advances in the NLP text domain. Although the survey provided a timely and deep perspective on both detection as well as mitigation of backdoor attacks, the authors fell short of developing a unified framework that integrates existing literature and exposes topics worth additional research. The authors also did not provide a complete treatment of the various backdoor attacks such as poisoning attacks, adversarial attacks, and the relationships among them.

In [40], Gao et al. developed a unified framework on the research advances, current trends, and challenges of backdoor attacks and defenses from the lenses of deep neural networks and provided a systematic and comprehensive view of this emerging research area. The authors argued that backdoor attack surfaces lack a systematic taxonomy according to the attacker's capabilities. As a result, attacks are diverse and not combed. Furthermore, the survey shows various nascent backdoor countermeasures have not yet been analyzed and compared. The latest trend to develop more efficient countermeasures is uneasy in this context. The authors provided the research community with an up-to-date review of backdoor attacks and countermeasures in the study. This particular research survey's unique property is that the authors developed four general categories of countermeasures: blind backdoor removal, offline backdoor inspection, online backdoor inspection, and post backdoor removal. Moreover, they reviewed countermeasures and exposed their advantages and disadvantages. Despite the unified and comprehensive taxonomy of backdoor attacks and countermeasures, authors stop short of extending their research effort to the NLP landscape and offer no insights on future research directions from the perspective of the text domain.

In [54], Gu et al. conducted an extensive research study on backdoor attacks and defenses spanning both the vision and video domains. In their survey, the authors provided a deep dive into the various backdoor attacks associated with the training process of deep neural networks and integrated existing literature on the mitigation and removal of backdoor trojans. The survey stated that the defense side of backdoor attacks lags far behind the attack side and that no single defense can mitigate most types of backdoor attacks. Hence, the authors offered insights into future research directions pertaining to developing more practical and general backdoor detection mechanisms to combat the growing security concerns associated with deploying ML models to various application domains. It's worth noting that this is the first research work to state the importance of extending research surveys to the NLP domain because, according to the authors, the research literature on survey works addressing backdoor attacks and defenses in the text domain is highly under-explored and call for further research. A gap our research survey intends to close [11, 31, 32, 37, 138].

Since research on backdoor attacks and defenses is developing rapidly, this survey intends to provide an overview and discussion of current and emerging techniques. Instead of summarizing only limited research [100] or analyzing existing methods simply based on adversary capabilities, we provide a brief but comprehensive overview as well as a taxonomy of existing methods. As far as we know, this is the first systematic taxonomy for backdoor attacks and defenses in the NLP domain. It will facilitate the design of more advanced methods by helping researchers and practitioners identify the properties and limitations of each method. Furthermore, our survey hopes to inspire a broader understanding of backdoor attacks and defenses to facilitate the design of more robust and secure NLP models [57].

## 8 CONCLUSION

This work extensively covers research efforts on backdoor learning for NLP. To this end, we systematically and comprehensively survey state-of-the-art research studies on backdoor attacks and defenses. Additionally, we thoroughly review and analyze various aspects of backdoor learning, including techniques, model architectures, evaluation metrics, and benchmark datasets. We argue that for backdoor learning to contribute to actual robustness, research studies should take into account an expansive view and strive to answer questions related to why such attacks and defenses are successful. It is crucial to determine whether any given technique is booming due to limitations and weaknesses associated with the target model (inherent incapability arising from intrinsic properties of a target model) or whether it is due to weaknesses or limitations in the dataset itself. Finally, we offer insights into open challenges and future research directions worth pursuing.

Table 1. A COMPARISON BETWEEN VARIOUS WORKS FROM THE LITERATURE, ACROSS TECHNIQUES, METRICS, BENCH-MARKS, ATTACKS, DEFENSES, THREAT MODEL, AND ATTACK GRANULARITY (WHITE-BOX VS BLACK-BOX), AND LINGUISTIC TASK

| Study | Year | Technique | Dataset | Linguistic Task | Threat Model | Model Architecture | |
|---|---|---|---|---|---|---|---|
| Yang et al.[181] | 2021 | Embedding Poisoning | IMDB,Amazon, Yelp, Twitter | sentiment analysis, toxic detection | white-box | BERT | |
| Liu et al. [98] | 2022 | Gradient-based Adversarial Attacks | Amazon, TrojanAI | sentiment analysis | generic | BERT, GPT, LSTM, GRU | |
| Azizi et al. [6] | 2021 | Sequence-to-sequence (seq-2-seq) generative model | MR, Yelp, AG News, HS | sentiment analysis, topic classification | black-box | BERT, CNN, LSTM | |
| Chan et al. [14] | 2020 | Conditional Adversarially Regularized Autoencoder | SNLI, Yelp, MNLI | sentiment analysis, NLI | generic | BERT, RoBERTa, XLNET | |
| Chan et al. [16] | 2021 | Backdoor sentence insertion | IMDB, DBPedia | SA, SC, PR | grey-box | LSTM | |
| Chan et al. [21] | 2021 | trigger construction | IMDB, SST-5 | SA | white-box | BERT,LSTM | |
| Eger et al. [35] | 2019 | Visual text perturbations | IMDB, MR | SA | black-box | BERT,LSTM | |
| Nguyen et al. [119] | 2020 | Input-aware trigger generator via diversity loss | MNIST, Object Recognition | SA, SC | ResNet,LSTM | Acc, ASR | |
| Qi et al. [131] | 2021 | Syntactic trigger-based attack | SSt-2, AG NEWS, OLID | SA, SC | white-box | BERT,LSTM | |
| Qi et al. [132] | 2021 | Invisible triggers via learnable combination of word substitution | SSt-2, AG NEWS, OLID | SA, SC | white-box | BERT,LSTM | |
| Wallace et al. [159] | 2019 | Gradient guided search over token | SNLI, SQuAD, OLID | NLI | black-box | BERT,LSTM | |
| Yang et al. [180] | 2021 | Poisoned word embeddings | SST-2, IMDB, SNLI | SA, SC, NLI | black-box | BERT | |
| Yang et al. [182] | 2021 | Negative data augmentation and modifying word embeddings | Yelp, IMDB, Twitter | SA, toxic detection | black-box | BERT | |
| Zhang et al. [192] | 2021 | Re-weighted training of language models | WebText, Twitter | toxic detection, QA | white-box | BERT, GPT-2, XLNet | |
| Zhang et al. [195] | 2021 | Neuron-level backdoor attack | OLID, GTSRB, SST-2, Enron | toxic and spam detection, SA | black-box | BERT, RoBERTa, VGGNet | |
| Li et al. [83] | 2021 | Knowledge distillation | GTSRB, CFAIR-10 | image recognition | black-box | Resnet, VGGNet | |
| Garg et al. [44] | 2020 | Backdoor Injection by Adversarial Weight Perturbation | MR, CFAIR-10 | Generic | black-box | WordCNN, LSTM | |
| Chen et al. [19] | 2021 | Task-agnostics label replacement foundation model | SST-2, QNLI, RTE | Generic | white-box | BERT, GPT-2 | |
| Gan et al. [39] | 2021 | Triggerless genetic clean-labels sentence generation | SST-2, AG NEWS, OLID | SC, SA | black-box | BERT | |

## 9 ACKNOWLEDGEMENT

I would like to thank my PhD advisor, Dr. Gita Sukthankar, for her thoughtful feedback and sharp insights on the manuscript.